\documentclass[showpacs]{revtex4} %twocolumn
\usepackage{epsfig,amsmath,amssymb,graphicx,graphics,upgreek,textcomp} %
\usepackage{natbib}
\usepackage[english]{babel} %

\begin{document}

\vspace{0mm}
\title{Statistical description of Fermi system over a surface in a
uniform external field} %
\author{Yu.M. Poluektov}
%\email{yuripoluektov@kipt.kharkov.ua (y.poluekt52@gmail.com)} %
\affiliation{National Science Center ``Kharkov Institute of Physics and Technology'', 61108 Kharkov, Ukraine} %
\author{A.A. Soroka} %
\affiliation{National Science Center ``Kharkov Institute of Physics and Technology'', 61108 Kharkov, Ukraine} %

\begin{abstract}
A statistical approach to the description of the thermodynamic
properties of the Fermi particle system occupying a half-space over
a plane of finite size in a uniform external field is proposed. The
number of particles per unit area is assumed to be arbitrary, in
particular, small. General formulas are obtained for entropy,
energy, thermodynamic potential, heat capacities under various
conditions and the distribution of the particle number density over
the surface. In the continuum limit of a large surface area, the
temperature dependences of heat capacities and density distribution
are calculated. The cases of gravitational and electric fields are
considered.
\newline%
\end{abstract}

\maketitle

\vspace{10mm}
\section{Introduction}\vspace{-0mm} %\cite{}
Currently, increasing attention is being paid to investigating the
quantum properties of systems with a small number of particles in
confined volumes, such as quantum dots, mesoscopic objects and
nanostructures. Therefore, the problem of describing the properties
of such objects, taking into account their interaction with the
external environment and in external fields, is of practical interest. %
Statistical description is usually used to study systems with a very
large number of particles. However, statistical methods can also be
applied to study the equilibrium states of systems with a small
number of particles and even a single particle. When considering a
system within a grand canonical ensemble, it is assumed that it is a
part of a very large system, a thermostat, with which it can
exchange energy and particles. The thermostat itself is
characterized by such statistical quantities as temperature $T$ and
chemical potential $\mu$. Assuming that the subsystem under
consideration is in thermodynamic equilibrium with the thermostat,
the subsystem itself, even consisting of a small number of
particles, is characterized by the same quantities. For example, one
can consider the thermodynamics of an individual quantum oscillator
\cite{LL}. In the case when particle exchange with the thermostat is
possible, the time-averaged number of particles of a small subsystem
may not be an integer and, in particular, even less than unity.

A phenomenological generalization of thermodynamics for an ensemble
of non-interacting small systems was previously proposed in
\cite{Hill}. In work \cite{PS1}, the authors of this article
obtained expressions for entropy and equations for quantum
distribution functions in systems of non-interacting fermions and
bosons with an arbitrary, and also small, number of particles. Using
the approach developed in \cite{PS1}, the temperature dependences of
entropy, heat capacities and pressure in the two-level Fermi and
Bose systems were calculated in work \cite{PS2}. %
%The case of low temperatures is considered, when the discrete
%structure of the energy spectrum arising due to the size effect becomes important. %
%The  results of \cite{PS3} and \cite{PS3b} differ quantitatively
%since the energy spectrum of levels and their degeneracy are
%different in the cubic cavity and the sphere.
In works \cite{PS3},\cite{PS3b} thermodynamics of the Fermis gas
inside a cubic cavity and a sphere, respectively, for arbitrary
temperatures, number of particles and system sizes was studied. %
At low temperatures and small number of particles, these systems are
shown to demonstrate rich thermodynamics due to the variety of ways
particles can occupy degenerate discrete energy levels with increasing temperature. %
In work \cite{GDMD}, the thermodynamic properties of quantum dots of
ellipsoidal, cylindrical, cubic, and pyramidal shapes were investigated. %

Of undoubted interest is the study of the influence of external
fields on the states of low-dimensional and small systems, since
with the help of external fields it is possible to change their
characteristics and thus to control their properties.

The purpose of this paper is to develop a method for the statistical
description of a system of an arbitrary, including small, number of
Fermi particles being in a constant uniform external field over a
flat surface of finite size. Using the proposed approach, the
thermodynamic characteristics of the system, in particular the heat
capacities under various conditions, are found. The general results
are applied to the description of Fermi particles in the continuum
limit at a large surface area for gravitational and electric fields.

\section{Fermi particle in a uniform field}\vspace{-0mm} %
Let us consider the states of a Fermi particle located over the
plane $z=0$, which has the form of a square with sides of length $L$
along the $x,y$ axes. The $z$ coordinate changes in the region $0\le
z<\infty$. We assume that the uniform field is directed along the
$z$ axis:
\begin{equation} \label{01}
\begin{array}{l}
\displaystyle{%
  U(z)=Fz, \qquad F>0. %
}%
\end{array}
\end{equation}
The solution of the Schr\"{o}dinger equation has the form
\begin{equation} \label{02}
\begin{array}{l}
\displaystyle{%
  \psi(x,y,z)=C\frac{2}{L}\sin\!\Big(2\pi n_x\frac{x}{L}\Big)\sin\!\Big(2\pi n_y\frac{y}{L}\Big)\,\psi(z), %
}%
\end{array}
\end{equation}
where $n_x,n_y=\pm 1, \pm 2, \ldots$, and
$C=\left(\int_0^\infty\psi^2(z)dz\right)^{-1\!/2}$ is the
normalization factor. It is assumed that the boundary condition
$\psi\big(\pm L/2, \pm L/2, z\big)=0$ is satisfied, and the function
$\psi(z)$ is equal to zero on the surface, tends to zero at
$z\rightarrow\infty$ and satisfies the equation
\begin{equation} \label{03}
\begin{array}{l}
\displaystyle{%
  \frac{d^2\psi(z)}{dz^2}+\frac{2m}{\hbar^2}\Big[E_{||}-Fz\Big]\psi(z)=0, %
}%
\end{array}
\end{equation}
$m$ -- the particle mass, $E_{||}$ -- the particle energy of motion
along the field. The total energy of a fermion is the sum of its
energy of motion in the $(x,y)$ plane and its energy of motion along
the field
\begin{equation} \label{04}
\begin{array}{l}
\displaystyle{%
  E=\frac{\hbar^2}{2m}\!\left(\frac{2\pi}{L}\right)^{\!\!2}\big(n_x^2+n_y^2\big)+E_{||}. %
}%
\end{array}
\end{equation}
It is convenient to introduce the characteristic length $l$ and the
dimensionless energy $\varepsilon$:
\begin{equation} \label{05}
\begin{array}{l}
\displaystyle{%
  \frac{2mE_{||}}{\hbar^2}=\frac{\varepsilon}{l^2}, \qquad %
  \frac{2mF}{\hbar^2}=\frac{1}{l^3}.
}%
\end{array}
\end{equation}
Then equation (\ref{03}) will take the form
\begin{equation} \label{06}
\begin{array}{l}
\displaystyle{%
  \frac{d^2\psi(\tilde{z})}{d\tilde{z}^2}-\big(\tilde{z}-\varepsilon\big)\psi(\tilde{z})=0, %
}%
\end{array}
\end{equation}
where $\tilde{z}\equiv z/l$ -- the dimensionless coordinate. The
solutions of equation (\ref{06}) are the Airy functions ${\rm
Ai}(\tilde{z}-\varepsilon), {\rm Bi}(\tilde{z}-\varepsilon)$
\cite{AS}. If $\tilde{z}-\varepsilon>0$, then the Airy functions are
expressed through the Bessel functions of imaginary argument
\begin{equation} \label{07}
\begin{array}{l}
\displaystyle{%
  {\rm Ai}(\tilde{z}-\varepsilon)=\frac{1}{3}\,\sqrt{\tilde{z}-\varepsilon}\,\big[I_{-1/3}(\zeta)-I_{1/3}(\zeta)\big] %
  =\pi^{-1}\sqrt{(\tilde{z}-\varepsilon)/3}\,\,K_{1/3}(\zeta),
}%
\end{array}
\end{equation} \vspace{-4mm}
\begin{equation} \label{08}
\begin{array}{l}
\displaystyle{%
   {\rm Bi}(\tilde{z}-\varepsilon)=\sqrt{(\tilde{z}-\varepsilon)/3}\,\big[I_{-1/3}(\zeta)+I_{1/3}(\zeta)\big],
}%
\end{array}
\end{equation}
where $\zeta\equiv(2/3)\big|\tilde{z}-\varepsilon\big|^{3/2}$. If
$\tilde{z}-\varepsilon<0$, then these functions are expressed
through the Bessel functions of real argument
\begin{equation} \label{09}
\begin{array}{l}
\displaystyle{%
  {\rm Ai}(\tilde{z}-\varepsilon)=\frac{1}{3}\,\sqrt{\varepsilon-\tilde{z}}\,\big[J_{-1/3}(\zeta)+J_{1/3}(\zeta)\big], %
}%
\end{array}
\end{equation} \vspace{-4mm}
\begin{equation} \label{10}
\begin{array}{l}
\displaystyle{%
   {\rm Bi}(\tilde{z}-\varepsilon)=\sqrt{(\varepsilon-\tilde{z})/3}\,\big[J_{-1/3}(\zeta)-J_{1/3}(\zeta)\big]. %
}%
\end{array}
\end{equation}
For the boundary conditions of the problem under consideration
$\psi(0)=0$, $\psi(\infty)=0$, the normalized wave functions have the form %
\begin{equation} \label{12}
\begin{array}{l}
\displaystyle{%
  \psi_n(\tilde{z})=\frac{1}{\sqrt{l}}\frac{{\rm Ai}(\tilde{z}-\varepsilon_n)}{{\rm Ai}'(-\varepsilon_n)}, %
}%
\end{array}
\end{equation}
where %
${\rm Ai}'(-\varepsilon_n\big)=-\frac{1}{3}\,\varepsilon_n %
\Big[J_{-2/3}\Big(\frac{2}{3}\,\varepsilon_n^{3/2}\Big)-J_{2/3}\Big(\frac{2}{3}\,\varepsilon_n^{3/2}\Big)\Big]$
\cite{AS}. The index $n=1,2,\ldots$ numbers the energy levels. To
find the asymptotics at infinity it should be taken into account
that $K_\nu(\zeta)\sim\sqrt{\frac{\pi}{2\zeta}}\,e^{-\zeta}$ at $\zeta\rightarrow\infty$, so that %
\begin{equation} \label{13}
\begin{array}{l}
\displaystyle{%
  \psi(\tilde{z})\sim\frac{1}{2\sqrt{\pi}\,{\rm Ai}'(-\varepsilon)}\,(\tilde{z}-\varepsilon)^{-1/4}\,e^{-\frac{2}{3}\,(\tilde{z}-\varepsilon)^{3/2}}. %
}%
\end{array}
\end{equation}
The energy levels are determined from the condition for the wave
function on the surface ${\rm Ai}(-\varepsilon_n)=0$:
\begin{equation} \label{14}
\begin{array}{l}
%\displaystyle{%
  J_{-1/3}\Big(\frac{2}{3}\,\varepsilon_n^{3/2}\Big)+J_{1/3}\Big(\frac{2}{3}\,\varepsilon_n^{3/2}\Big)=0. %
%}%
\end{array}
\end{equation}
To determine the energy of high levels $\varepsilon_n\gg 1$, one can
use the asymptotics
\begin{equation} \label{15}
\begin{array}{l}
\displaystyle{%
  J_{1/3}(\zeta)\sim\sqrt{\frac{2}{\pi\zeta}}\,\cos\!\Big(\zeta-\frac{5\pi}{12}\Big), \qquad%
  J_{-1/3}(\zeta)\sim\sqrt{\frac{2}{\pi\zeta}}\,\cos\!\left(\zeta-\frac{\pi}{12}\right). %
}%
\end{array}
\end{equation}
From here we have the asymptotics for the wave function at $\varepsilon_n-\tilde{z}\gg 1$: %
\begin{equation} \label{16}
\begin{array}{l}
\displaystyle{%
  \psi(\tilde{z})\sim\frac{1}{\sqrt{\pi}\,{\rm Ai}'(-\varepsilon_n)}\,(\varepsilon_n-\tilde{z})^{-1/4} %
  \cos\!\Big(\,\frac{2}{3}(\varepsilon_n-\tilde{z})^{3/2}-\frac{\pi}{4}\,\Big). %
}%
\end{array}
\end{equation}
Then from the condition %
$\cos\!\Big(\frac{2}{3}\,\varepsilon_n^{3/2}-\frac{\pi}{4}\Big)=0$
we find the formula for the energy spectrum of high levels
\begin{equation} \label{17}
\begin{array}{l}
\displaystyle{%
  \varepsilon_n=\bigg[\frac{3\pi}{2}\bigg(n-\frac{1}{4}\bigg)\bigg]^{\!2/3}. %
}%
\end{array}
\end{equation}
Note that the approximate formula (\ref{17}) gives values close to
the exact result. Even for the first level, the calculation using
the exact formula gives $\varepsilon_1=2.338$, while using formula
(\ref{17}) $\varepsilon_1=2.320$. For higher levels, as can be seen
from Table\,I, the accuracy increases, so that in the case of the
half-space formula (\ref{17}) can be used almost always.
\vspace{-4mm}%
\begin{table}[h!] \nonumber
\centering %
\caption{Discrete levels of motion along the field in the order of increasing energy} %
\vspace{0.5mm}%
\begin{tabular}{|c|c|c|c|c|c|c|c|c|c|c|} \hline  %
\rule{0mm}{0pt} $\varepsilon_n$             \rule{0mm}{0pt} & \rule{0mm}{0pt} 2.338 \rule{0mm}{0pt}  & \rule{0mm}{0pt} 4.088 \rule{0mm}{0pt}  & \rule{0mm}{0pt} 5.521 \rule{0mm}{0pt}  & \rule{0mm}{0pt} 6.787 \rule{0mm}{0pt} & \rule{0mm}{0pt} 7.944 \rule{0mm}{0pt} & \rule{0mm}{0pt} 9.023 \rule{0mm}{0pt} & \rule{0mm}{0pt} 10.040 \rule{0mm}{0pt} & \rule{0mm}{0pt} 11.009 \rule{0mm}{0pt}& \rule{0mm}{0pt} 11.936 \rule{0mm}{0pt} & \rule{0mm}{0pt} 12.829 \rule{0mm}{0pt} \\ \hline %
\rule{0mm}{0pt} $\varepsilon_n$\,(\ref{17}) \rule{0mm}{0pt} & \rule{0mm}{0pt} 2.320 \rule{0mm}{0pt}  & \rule{0mm}{0pt} 4.082 \rule{0mm}{0pt}  & \rule{0mm}{0pt} 5.517 \rule{0mm}{0pt}  & \rule{0mm}{0pt} 6.785 \rule{0mm}{0pt} & \rule{0mm}{0pt} 7.943 \rule{0mm}{0pt} & \rule{0mm}{0pt} 9.021 \rule{0mm}{0pt} & \rule{0mm}{0pt} 10.039 \rule{0mm}{0pt} & \rule{0mm}{0pt} 11.008 \rule{0mm}{0pt}& \rule{0mm}{0pt} 11.935 \rule{0mm}{0pt} & \rule{0mm}{0pt} 12.828 \rule{0mm}{0pt} \\ \hline %
\end{tabular}  %tabular
\vspace{0mm}
\end{table}
\vspace{0mm}%

The form of the wave functions for the first three levels is shown in Fig.\,1. %
\begin{figure}[h!]
\vspace{-2mm}  \hspace{0mm}
\includegraphics[width = 7.44cm]{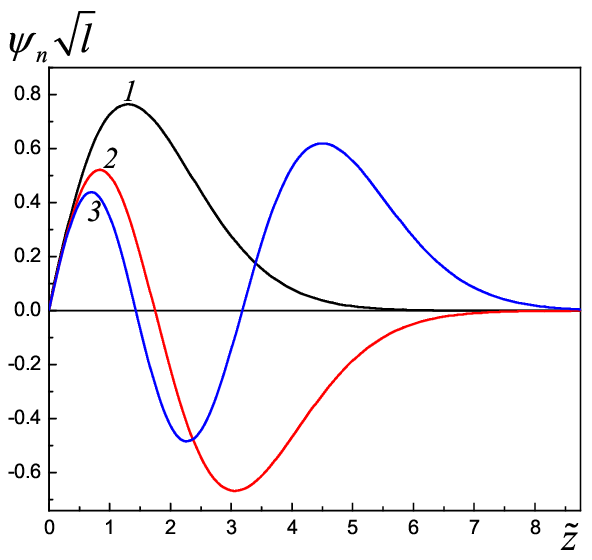} % 1.0\columnwidth
\vspace{-4mm} %
\caption{\label{fig01} %\hspace{10mm} %
Wave functions $\psi_n(\tilde{z})\sqrt{l}$\, for the first three
levels: ({\it 1}) $n=1$, ({\it 2}) $n=2$, ({\it 3}) $n=3$. %
}%
\end{figure}

The full energy spectrum of the fermion (\ref{04}) can be written as
\begin{equation} \label{18}
\begin{array}{l}
\displaystyle{%
  E_{(i,n)}=\frac{\hbar^2}{2m}\left(\frac{2\pi}{L}\right)^{\!\!2}\gamma_i^2 + \frac{\hbar^2}{2ml^2}\,\varepsilon_n, %
}%
\end{array}
\end{equation}
where $\gamma_i^2\equiv n_x^2+n_y^2$. Here the first term determines
the contribution to the energy of motion in the $(x,y)$ plane, and
the second term determines the contribution of motion along the
field along the $z$ axis. The discrete energy levels $\varepsilon_n$
are not degenerate, and the degeneracy factor of a level with a
given $\gamma_i$ with account of the two-fold degeneracy in the spin
projection will be denoted by $r_i$. The first ten bottom values of
the parameter $\gamma_i^2$ in the order of increasing energy and the
level degeneracy factor $r_i$ are given in Table\,II.
\newpage
\begin{table}[h!] \nonumber
\begin{center}
\caption{\mbox{}{The first ten values of the parameter $\gamma_i^2$
that determine the energy of motion in the $(x,y)$ plane and the
degeneracy factor $r_i$ with account of the two-fold degeneracy in
the spin projection}}
\vspace{0.5mm}%
\begin{tabular}{|c|c|c|c|c|c|c|c|c|c|c|} \hline  % \rule{3mm}{0pt} 0.385 \rule{3mm}{0pt}
\rule{3mm}{0pt} $i$          \rule{3mm}{0pt} & \rule{3mm}{0pt} 1 \rule{3mm}{0pt}  & \rule{3mm}{0pt} 2  \rule{3mm}{0pt}  & \rule{3mm}{0pt} 3 \rule{3mm}{0pt}  & \rule{3mm}{0pt} 4 \rule{3mm}{0pt}  & \rule{3mm}{0pt} 5 \rule{3mm}{0pt}  & \rule{3mm}{0pt} 6 \rule{3mm}{0pt} & \rule{3mm}{0pt} 7  \rule{3mm}{0pt} & \rule{3mm}{0pt} 8  \rule{3mm}{0pt}& \rule{3mm}{0pt} 9  \rule{3mm}{0pt}& \rule{3mm}{0pt} 10 \rule{3mm}{0pt} \\ \hline %
\rule{3mm}{0pt} $\gamma_i^2$ \rule{3mm}{0pt} & \rule{3mm}{0pt} 2 \rule{3mm}{0pt}  & \rule{3mm}{0pt} 5  \rule{3mm}{0pt}  & \rule{3mm}{0pt} 8 \rule{3mm}{0pt}  & \rule{3mm}{0pt} 10 \rule{3mm}{0pt} & \rule{3mm}{0pt} 13 \rule{3mm}{0pt} & \rule{3mm}{0pt} 17 \rule{3mm}{0pt}& \rule{3mm}{0pt} 18 \rule{3mm}{0pt} & \rule{3mm}{0pt} 20 \rule{3mm}{0pt}& \rule{3mm}{0pt} 25 \rule{3mm}{0pt}& \rule{3mm}{0pt} 26 \rule{3mm}{0pt} \\ \hline %
\rule{3mm}{0pt} $r_i$        \rule{3mm}{0pt} & \rule{3mm}{0pt} 8 \rule{3mm}{0pt}  & \rule{3mm}{0pt} 16 \rule{3mm}{0pt} & \rule{3mm}{0pt}  8 \rule{3mm}{0pt} & \rule{3mm}{0pt} 16 \rule{3mm}{0pt}  & \rule{3mm}{0pt} 16 \rule{3mm}{0pt} & \rule{3mm}{0pt} 16 \rule{3mm}{0pt}& \rule{3mm}{0pt}  8 \rule{3mm}{0pt} & \rule{3mm}{0pt} 16 \rule{3mm}{0pt}& \rule{3mm}{0pt} 16 \rule{3mm}{0pt}& \rule{3mm}{0pt} 16 \rule{3mm}{0pt} \\ \hline %
\end{tabular}
\end{center}
\vspace{-5mm}
\end{table}
\noindent Using the formulas given in this section, we formulate a
statistical description of the Fermi system of an arbitrary number
of particles over a flat surface, without assuming in advance that
the area $A\equiv L^2$ is large and without passing to the
thermodynamic limit $L\rightarrow\infty$. The proposed approach is
also applicable to the description of size effects caused by a
finite value of the area.

\section{Statistical description of Fermi particles in a uniform field} \vspace{-1mm}%
When constructing thermodynamics on the basis of the statistical
method, we will proceed from the formula for entropy $S=\sum_\nu S_\nu$: %
\begin{equation} \label{19}
\begin{array}{l}
\displaystyle{%
   S_\nu = \ln\Gamma(r_i+1) -\ln\Gamma(r_if_\nu +1)-\ln\Gamma\big[r_i(1-f_\nu)+1\big], %
}
\end{array}
\end{equation}
where $r_i$ -- the degeneracy factor of a level, $\nu\equiv (i,n)$,
and $f_\nu\equiv f_{(i,n)}$ -- the population of a level. We used
the definition of factorial through the gamma function $N!=\Gamma(N+1)$. %
This makes it possible to study systems in which the time-averaged
number of particles is not large and integer and to consider
$0<N<\infty$ as a continuous positive parameter \cite{PS1,PS2,PS3,PS3b}. %
The total energy of the whole system
\begin{equation} \label{20}
\begin{array}{l}
\displaystyle{%
   E=\sum_\nu E_\nu f_\nu r_i, %
}
\end{array}
\end{equation}
where $E_\nu\equiv E_{(i,n)}$ is given by formula (\ref{18}), and
the total number of particles
\begin{equation} \label{21}
\begin{array}{l}
\displaystyle{%
   N=\sum_\nu f_\nu r_i. %
}
\end{array}
\end{equation}
The average number of particles $f_\nu=N_\nu/r_i$ at level $\nu$, or
the population of the level, is found from the condition
\begin{equation} \label{22}
\begin{array}{l}
\displaystyle{%
   \frac{\partial}{\partial f_\nu}\big(S-\alpha N-\beta E\big)=0,  %
}
\end{array}
\end{equation}
where $\alpha=1/T$, $\beta=\mu/T$ are Lagrange multipliers. From
here we find the equation that determines the average number of
particles at level $\nu$
\begin{equation} \label{23}
\begin{array}{l}
\displaystyle{%
   \theta_\nu\equiv\theta(f_\nu,r_i)\equiv\psi\big[r_i(1-f_\nu)+1\big] -\psi\big(r_if_\nu+1\big)=\frac{(E_\nu-\mu)}{T},  %
}
\end{array}
\end{equation}
where $T$ -- temperature, $\mu$ -- chemical potential,
$\psi(x)=d\ln\Gamma(x)\big/dx$ -- the logarithmic derivative of the
gamma function (the psi function) \cite{AS}.

The thermodynamic potential is defined by the usual formula
$\Omega=E-TS-\mu N$. The differential of the thermodynamic potential has the form %
\begin{equation} \label{24}
\begin{array}{l}
\displaystyle{%
  d\Omega=\sum_\nu r_i f_\nu\,dE_\nu -SdT-Nd\mu. %
}
\end{array}
\end{equation}
Taking into account formula (\ref{18}) and $dl/l=-dF\big/3F$, we find %
\begin{equation} \label{25}
\begin{array}{l}
\displaystyle{%
  dE_\nu=-\frac{\hbar^2}{2m}\left(\frac{2\pi}{A}\right)^{\!\!2}\gamma_i^2\,dA + \frac{2l}{3}\,\varepsilon_n\,dF. %
}%
\end{array}
\end{equation}
Therefore, the differential (\ref{24}) can be represented in the form %
\begin{equation} \label{26}
\begin{array}{l}
\displaystyle{%
  d\Omega=-SdT-Nd\mu + \sigma dA + D dF. %
}
\end{array}
\end{equation}
Here the quantity
\begin{equation} \label{27}
\begin{array}{l}
\displaystyle{%
   \sigma=\bigg(\frac{\partial \Omega}{\partial A}\bigg)_{T,\mu,F}= %
   -\frac{\hbar^2}{2m}\left(\frac{2\pi}{A}\right)^{\!\!2}\sum_\nu r_i f_\nu\gamma_i^2 %
}%
\end{array}
\end{equation}
has the meaning of the surface tension, and the quantity
\begin{equation} \label{28}
\begin{array}{l}
\displaystyle{%
   D=\bigg(\frac{\partial \Omega}{\partial F}\bigg)_{T,\mu,A}= %
   \frac{2l}{3}\sum_\nu r_i f_\nu \varepsilon_n %
}%
\end{array}
\end{equation}
can naturally be called ``induction''.

To calculate heat capacities and other thermodynamic coefficients,
let us first present the following differentials
\begin{equation} \label{29}
\begin{array}{l}
\displaystyle{%
   r_idf_\nu=\pi\!\left(\frac{\Lambda}{A}\right)^{\!\!2}\frac{\gamma_i^2}{\theta_\nu^{(1)}}\,dA\, - %
   \frac{2l}{3T\theta_\nu^{(1)}}\,\varepsilon_n dF\, + \frac{d\mu}{\theta_\nu^{(1)}T} + \frac{\theta_\nu}{\theta_\nu^{(1)}}\frac{dT}{T},  %
}%
\end{array}
\end{equation} \vspace{-2mm}
\begin{equation} \label{30}
\begin{array}{l}
\displaystyle{%
   dN=\pi\!\left(\frac{\Lambda}{A}\right)^{\!\!2}\chi_3dA\, - %
   \frac{2l}{3T}\,\chi_6 dF\, + \frac{1}{\theta^{(1)}}\frac{d\mu}{T} + \chi_1\frac{dT}{T},  %
}%
\end{array}
\end{equation} \vspace{-2mm}
\begin{equation} \label{31}
\begin{array}{l}
\displaystyle{%
   dS=\pi\!\left(\frac{\Lambda}{A}\right)^{\!\!2}\chi_4dA\, - %
   \frac{2l}{3T}\,\chi_7 dF\, + \chi_1\frac{d\mu}{T} + \chi_2\frac{dT}{T},  %
}%
\end{array}
\end{equation} \vspace{-2mm}
\begin{equation} \label{32}
\begin{array}{l}
\displaystyle{%
   -d\sigma\frac{1}{\pi T}\left(\frac{A}{\Lambda}\right)^{\!\!2} = %
   \bigg[-\frac{2}{A}\,\chi_{10} + \pi\!\left(\frac{\Lambda}{A}\right)^{\!\!2}\chi_5\bigg]dA - %
   \frac{2l}{3T}\,\chi_8 dF\, + \chi_3\frac{d\mu}{T} + \chi_4\frac{dT}{T},  %
}%
\end{array}
\end{equation} \vspace{-2mm}
\begin{equation} \label{33}
\begin{array}{l}
\displaystyle{%
  dD = \frac{2l}{3}\,\pi\!\left(\frac{\Lambda}{A}\right)^{\!\!2}\chi_8 dA -  %
  \frac{2l}{9F}\bigg(\chi_{11}+2\,\frac{lF}{T}\,\chi_9\bigg)dF + %
  \frac{2l}{3}\,\chi_6 \frac{d\mu}{T} + \frac{2l}{3}\,\chi_7\frac{dT}{T}.  %
}%
\end{array}
\end{equation}
Here the de Broglie thermal wavelength
\begin{equation} \label{34}
\begin{array}{c}
\displaystyle{\hspace{0mm}%
  \Lambda\equiv\bigg(\frac{2\pi\hbar^2}{mT}\bigg)^{\!\!1/2} %
}%
\end{array}
\end{equation}
is introduced and the following notations are used
\begin{equation} \label{35}
\begin{array}{l}
\displaystyle{%
   \frac{1}{\theta^{(1)}}\equiv\sum_\nu\frac{1}{\theta_\nu^{(1)}}, \quad %
   \chi_1\equiv\sum_\nu\frac{\theta_\nu}{\theta_\nu^{(1)}}, \quad %
   \chi_2\equiv\sum_\nu\frac{\theta_\nu^2}{\theta_\nu^{(1)}}, \quad %
   \chi_3\equiv\sum_\nu\frac{\gamma_i^2}{\theta_\nu^{(1)}}, %
}\vspace{3mm}\\ %
\displaystyle{\hspace{0mm}%
   \chi_4\equiv\sum_\nu\frac{\theta_\nu}{\theta_\nu^{(1)}}\,\gamma_i^2, \quad %
   \chi_5\equiv\sum_\nu\frac{\gamma_i^4}{\theta_\nu^{(1)}}, \quad %
   \chi_6\equiv\sum_\nu\frac{\varepsilon_n}{\theta_\nu^{(1)}}, \quad %
   \chi_7\equiv\sum_\nu\frac{\theta_\nu}{\theta_\nu^{(1)}}\,\varepsilon_n, %
}\vspace{3mm}\\ %
\displaystyle{\hspace{0mm}%
   \chi_8\equiv\sum_\nu\frac{\varepsilon_n\gamma_i^2}{\theta_\nu^{(1)}}, \quad %
   \chi_9\equiv\sum_\nu\frac{\varepsilon_n^2}{\theta_\nu^{(1)}}, \quad %
   \chi_{10}\equiv\sum_\nu r_i f_\nu \gamma_i^2, \quad %
   \chi_{11}\equiv\sum_\nu r_i f_\nu\, \varepsilon_n,  %
}%
\end{array}
\end{equation}
where
\begin{equation} \label{36}
\begin{array}{l}
\displaystyle{%
   \theta_\nu^{(1)}\equiv\theta^{(1)}(f_\nu, r_i)\equiv\psi^{(1)}\big[r_i(1-f_\nu)+1\big] + \psi^{(1)}\big(r_if_\nu+1\big), %
}
\end{array}
\end{equation}
$\psi^{(1)}(x)=d\psi(x)/dx= d^{\,2}\ln\Gamma(x)\big/dx^2$ -- the trigamma function \cite{AS}. %

Usually there are considered systems with a fixed total number of
particles, when $dN=0$. In this case, the differential of chemical
potential can be expressed through the differentials of temperature,
area and field
\begin{equation} \label{37}
\begin{array}{l}
\displaystyle{%
  \frac{d\mu}{T}=-\pi\theta^{(1)}\left(\frac{\Lambda}{A}\right)^{\!\!2}\chi_3dA + %
  \theta^{(1)}\frac{2l}{3T}\,\chi_6 dF\, - \theta^{(1)}\chi_1\frac{dT}{T}.  %
}%
\end{array}
\end{equation}
Then the differentials (\ref{31})\,--\,(\ref{33}) will take the form
\begin{equation} \label{38}
\begin{array}{l}
\displaystyle{%
  dS=\pi\!\left(\frac{\Lambda}{A}\right)^{\!\!2}\eta_2 dA - %
  \frac{2l}{3T}\,\eta_4 dF\, +\eta_1\frac{dT}{T},  %
}%
\end{array}
\end{equation}\vspace{-4mm}
\begin{equation} \label{39}
\begin{array}{l}
\displaystyle{%
   -d\sigma\frac{1}{\pi T}\left(\frac{A}{\Lambda}\right)^{\!\!2} = %
   \bigg[-\frac{2}{A}\,\chi_{10} + \pi\!\left(\frac{\Lambda}{A}\right)^{\!\!2}\eta_3\bigg]dA - %
   \frac{2l}{3T}\,\eta_5 dF\, + \eta_2\frac{dT}{T},  %
}%
\end{array}
\end{equation}\vspace{-3mm}
\begin{equation} \label{40}
\begin{array}{l}
\displaystyle{%
  dD = \frac{2l}{3}\,\pi\!\left(\frac{\Lambda}{A}\right)^{\!\!2}\eta_5 dA -  %
  \frac{2l}{9F}\bigg(\chi_{11}+2\,\frac{lF}{T}\,\eta_6\bigg)dF + %
  \frac{2l}{3}\,\eta_4\frac{dT}{T}.  %
}%
\end{array}
\end{equation}
Here we used the notations
\begin{equation} \label{41}
\begin{array}{l}
\displaystyle{%
  \eta_1\equiv\chi_2-\theta^{(1)}\chi_1^2, \qquad \eta_2\equiv\chi_4-\theta^{(1)}\chi_1\chi_3, \qquad \eta_3\equiv\chi_5-\theta^{(1)}\chi_3^2, %
}\vspace{3mm}\\ %
\displaystyle{\hspace{0mm}%
  \eta_4\equiv\chi_7-\theta^{(1)}\chi_1\chi_6, \qquad \eta_5\equiv\chi_8-\theta^{(1)}\chi_3\chi_6, \qquad \eta_6\equiv\chi_9-\theta^{(1)}\chi_6^2. %
}%
\end{array}
\end{equation}

The obtained formulas allow to calculate heat capacities under
various conditions. The heat capacity under arbitrary conditions is
defined by the relation
\begin{equation} \label{42}
\begin{array}{l}
\displaystyle{%
  C=T\frac{dS}{dT}=\pi\!\left(\frac{\Lambda}{A}\right)^{\!\!2}\eta_2T\frac{dA}{dT} - %
  \frac{2l}{3}\,\eta_4\frac{dF}{dT}\, +\eta_1.  %
}%
\end{array}
\end{equation}
For a fixed area $dA=0$ and a constant field $dF=0$ we obviously have %
\begin{equation} \label{43}
\begin{array}{l}
\displaystyle{%
  C_{N,A,F}= \eta_1.  %
}%
\end{array}
\end{equation}
In the case of fixed surface tension $d\sigma=0$ and field $dF=0$,
from (\ref{39}) it follows
\begin{equation} \label{44}
\begin{array}{l}
\displaystyle{%
  T\frac{dA}{dT}=\Bigg[\frac{2}{A}\,\chi_{10} - \pi\!\left(\frac{\Lambda}{A}\right)^{\!\!2}\eta_3\Bigg]^{\!-1}\eta_2. %
}%
\end{array}
\end{equation}
With account of (\ref{44}) we find
\begin{equation} \label{45}
\begin{array}{l}
\displaystyle{%
  C_{N,F,\sigma}=\pi\!\left(\frac{\Lambda}{A}\right)^{\!\!2}\eta_2^2 %
  \Bigg[\frac{2}{A}\,\chi_{10} - \pi\!\left(\frac{\Lambda}{A}\right)^{\!\!2}\eta_3\Bigg]^{\!-1}+\eta_1. %
}%
\end{array}
\end{equation}
At fixed field $dF=0$ and induction $dD=0$, from (\ref{40}) it follows %
\begin{equation} \label{46}
\begin{array}{l}
\displaystyle{%
  T\frac{dA}{dT}=-\frac{1}{\pi}\!\left(\frac{A}{\Lambda}\right)^{\!\!2}\frac{\eta_4}{\eta_5}, %
}%
\end{array}
\end{equation}
so that
\begin{equation} \label{47}
\begin{array}{l}
\displaystyle{%
  C_{N,F,D}= -\frac{\eta_4}{\eta_5}\,\eta_2 + \eta_1.  %
}%
\end{array}
\end{equation}

The obtained general relations (\ref{43}),\,(\ref{45}),\,(\ref{47})
can be transformed with taking into account equation (\ref{23}),
which we represent in the form
\begin{equation} \label{48}
\begin{array}{l}
\displaystyle{%
  \theta_\nu=\pi l_L^2\gamma_i^2 + \frac{\varepsilon_n}{4\pi}\,l_F^2 - t,   %
}%
\end{array}
\end{equation}
where $t\equiv\mu/T$, and the ratios of the de Broglie thermal
wavelength (\ref{34}) to the characteristic lengths $L, l$ are introduced %
\begin{equation} \label{49}
\begin{array}{l}
\displaystyle{%
  l_L\equiv\frac{\Lambda}{L}, \qquad l_F\equiv\frac{\Lambda}{l}.   %
}%
\end{array}
\end{equation}
Given (\ref{48}), we find that the parameters
$\chi_1,\chi_2,\chi_4,\chi_7$ are expressed through the six
parameters $\chi_3,\chi_5,\chi_6,\chi_8,\chi_9,\theta^{(1)}$: %
\begin{equation} \label{50}
\begin{array}{c}
\displaystyle{%
  \chi_1=\pi l_L^2\chi_3 + \frac{l_F^2}{4\pi}\,\chi_6 - \frac{t}{\theta^{(1)}}, %
}\vspace{3mm}\\ %
\displaystyle{\hspace{0mm}%
  \chi_2=\pi^2 l_L^4\chi_5 + \frac{l_F^4}{(4\pi)^2}\,\chi_9 + \frac{t^2}{\theta^{(1)}}  %
  +\frac{1}{2}l_L^2l_F^2\chi_8 - 2\pi t\l_L^2\chi_3 - t\frac{l_F^2}{2\pi}\chi_6, %
}\vspace{3mm}\\ %
\displaystyle{\hspace{0mm}%
  \chi_4=\pi l_L^2\chi_5 + \frac{l_F^2}{4\pi}\,\chi_8 - t\chi_3, %
}\vspace{3mm}\\ %
\displaystyle{\hspace{0mm}%
  \chi_7=\pi l_L^2\chi_8 + \frac{l_F^2}{4\pi}\,\chi_9 - t\chi_6. %
}%
\end{array}
\end{equation}
Taking into account relations (\ref{50}), we find
\begin{equation} \label{51}
\begin{array}{c}
\displaystyle{%
  \eta_1=\pi^2 l_L^4\eta_3 + \frac{l_F^4}{(4\pi)^2}\,\eta_6 + \frac{1}{2}\,l_L^2l_F^2\eta_5, %
}\vspace{3mm}\\ %
\displaystyle{\hspace{0mm}%
  \eta_2=\pi l_L^2\eta_3 + \frac{l_F^2}{4\pi}\eta_5, \quad   \eta_4=\pi l_L^2\eta_5 + \frac{l_F^2}{4\pi}\eta_6. %
}%
\end{array}
\end{equation}
Thus, of the six quantities (\ref{41}) only three quantities
$\eta_3,\eta_5,\eta_6$ are independent. As a result, the heat
capacities (\ref{43}),\,(\ref{45}),\,(\ref{47}) take the form
\begin{equation} \label{52}
\begin{array}{c}
\displaystyle{%
  C_{N,A,F}=\pi^2 l_L^4\eta_3 + \frac{l_F^4}{(4\pi)^2}\,\eta_6 + \frac{1}{2}\,l_L^2l_F^2\eta_5, %
}%
\end{array}
\end{equation}\vspace{-4mm}
\begin{equation} \label{53}
\begin{array}{c}
\displaystyle{%
  C_{N,F,\sigma}=\pi\!\left(\frac{\Lambda}{A}\right)^{\!\!2}\!\bigg(\pi l_L^2\eta_3 + \frac{l_F^2}{4\pi}\eta_5\bigg)^{\!2}
  \bigg[\frac{2}{A}\,\chi_{10}-\pi\!\left(\frac{\Lambda}{A}\right)^{\!\!2}\eta_3\bigg]^{\!-1} + %
  \pi^2 l_L^4\eta_3 + \frac{l_F^4}{(4\pi)^2}\,\eta_6 + \frac{1}{2}\,l_L^2l_F^2\eta_5, %
}%
\end{array}
\end{equation}\vspace{-4mm}
\begin{equation} \label{54}
\begin{array}{c}
\displaystyle{%
  C_{N,F,D}=\frac{1}{4\eta_5}\,l_L^2l_F^2\big(\eta_5^2 - \eta_3\eta_6 \big). %
}%
\end{array}
\end{equation}

Obviously, the system under consideration is spatially inhomogeneous
in the direction of the field. The spatial dependence of the density
is determined by the spatial dependence of the square of the wave function %
\begin{equation} \label{55}
\begin{array}{c}
\displaystyle{%
  n(z)=\frac{1}{L^2}\sum_\nu\psi_n^2(z)f_\nu r_i = \frac{1}{L^2}\sum_n\psi_n^2(z)N_n,  %
}%
\end{array}
\end{equation}
where $N_n=\sum_i f_\nu r_i$ -- the total number of particles with
quantum number $(n)$, and the number of particles per unit area is
$n_A\equiv N/A=\int_0^\infty dz\,n(z) = A^{-1}\sum_nN_n$.

\section{Continual approximation} %
In the general formulas for thermodynamic quantities given in the
previous section, no restrictions on the size of the area $A=L^2$
were imposed, and the discrete structure of levels by quantum
numbers $(i,n)$ (\ref{18}) was taken into account. Thus, these
formulas are suitable for description of systems of arbitrary sizes
with an arbitrary number of particles and for study of size effects.
As the area $A$ increases, the distance between adjacent levels with
$\gamma_i^2$ and $\gamma_{i+1}^2$ decreases, so that in the limit
$L\rightarrow\infty$ we can pass to a continual description. At
that, the motion along the field remains quantized.

Since in the space of numbers $(n_x,n_y)$ there is a unit square per
one state, then the total number of states in a large system, for
which the condition $n_x^2+n_y^2<\gamma^2$ is satisfied, is equal to
the area of the circle $S(\gamma)=\pi\gamma^2$. The number of states
in the interval ($\gamma,\gamma + \Delta\gamma$) is $\Delta
S(\gamma)=2\pi\gamma\Delta\gamma$, so that the density of the number
of states on a circle of radius $\gamma$ is equal to the length of
the circle $s(\gamma)=2\pi\gamma$. The degeneracy factor of the
level with account of the two-fold degeneracy in the spin projection
is $r_i=4\pi\gamma_i$.

Let us obtain a formula for the number of particles in the continual
approximation. The number of particles with fixed $n$ and arbitrary
$i$ is $N_n=\sum_i r_if_{in}$. First assume that the number of
levels $M$ is finite and denote $k_j=(2\pi/L)\gamma_j$. Then the
total number of particles at level $(n)$ is
\begin{equation} \label{56}
\begin{array}{c}
\displaystyle{%
  N_n=2L\sum_{i=1}^M k_i f_{in}.  %
}%
\end{array}
\end{equation}
Now divide the interval of change of $k_j$ into equal intervals
$\Delta k\equiv (k_M-k_1)/(\gamma_M-\gamma_1)=2\pi/L$ and, taking
into account the definition of the de Broglie thermal wavelength
(\ref{34}), introduce dimensionless quantities
$\Delta\kappa\equiv\Lambda\Delta k=2\pi(\Lambda/L)$, and also
$\kappa_j=\Lambda k_j$. Then formula (\ref{56}) will take the form
\begin{equation} \label{57}
\begin{array}{c}
\displaystyle{%
  N_n=\frac{1}{\pi}\!\left(\frac{L}{\Lambda}\right)^{\!\!2}\sum_{i=1}^M f_{in}\kappa_i\Delta\kappa.  %
}%
\end{array}
\end{equation}
Let us consider the case $\Delta\kappa=2\pi(\Lambda/L)\ll 1$. In the
limit $L\rightarrow\infty$ this condition is true at any
temperature. In the case of a system with finite area $A$, this
condition is satisfied if $\Lambda\ll L/2\pi$ or $T^{1\!/2}\gg
\frac{2\pi\hbar}{L}\left(\frac{2\pi}{m}\right)^{\!1\!/2}$. At a
finite area one can pass to a continual description at high
temperatures, when the de Broglie thermal wavelength is much smaller
than the side $L$ of the square. Setting $M\rightarrow\infty$ and
passing from summation to integration in (\ref{57}), we obtain
\begin{equation} \label{58}
\begin{array}{c}
\displaystyle{%
  N_n=\frac{1}{\pi}\!\left(\frac{L}{\Lambda}\right)^{\!\!2}\int_0^\infty\!f_n\Big(\frac{\kappa}{\Lambda}\Big)\kappa d\kappa = %
  \frac{L^2}{\pi}\int_0^\infty\!f_n(k)kdk.
}%
\end{array}
\end{equation}

\begin{figure}[b!]
\vspace{-2mm}  \hspace{0mm}
\includegraphics[width = 7.35cm]{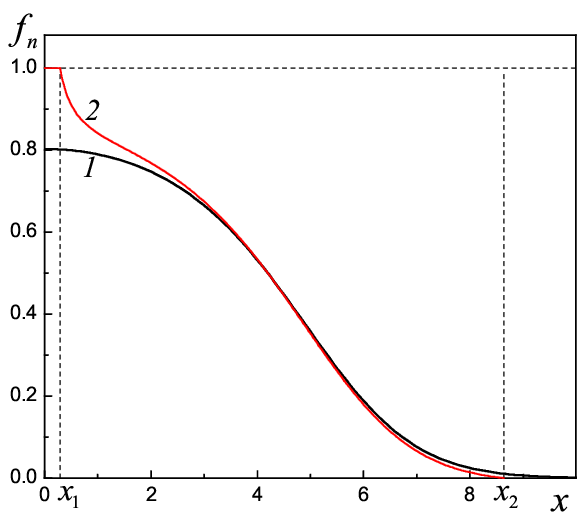} % 1.0\columnwidth
\vspace{-4mm} %
\caption{\label{fig02} %\hspace{10mm} %
The distribution functions: ({\it 1}) Fermi-Dirac $f_n^{{\rm FD}}(x;t_n)$ (\ref{61}), %
({\it 2}) in the continual approximation $f_n(x;t_n)$ by formula (\ref{60}), for $t_n=1.4$, $L/\Lambda=3$; %
$x\equiv k\Lambda$, $x_1=k_1\Lambda=0.29$, $x_2=k_2\Lambda=8.64$. %
}%
\end{figure}

The equation determining the average number of particles in each
state (\ref{23}) in this case can be written as
\begin{equation} \label{59}
\begin{array}{l}
\displaystyle{%
   \psi\big[r_j(1-f_n(k_j,t))+1\big] -\psi\big[r_jf_n(k_j,t)+1\big] = %
   \frac{1}{4\pi}\bigg[\big(\Lambda k_j\big)^2 + \left(\frac{\Lambda}{l}\right)^{\!\!2}\varepsilon_n\bigg] - t,  %
}
\end{array}
\end{equation}
where $t\equiv\mu/T$. In the continual approximation $k_j\rightarrow
k$ can be considered as a continuous variable, so that \linebreak
$r_i=2Lk_i\rightarrow 2Lk$. Equation (\ref{59}) in the continual
approximation takes the form
\begin{equation} \label{60}
\begin{array}{l}
\displaystyle{%
   \psi\big[2Lk(1-f_n(k,t))+1\big] -\psi\big[2Lk\,f_n(k,t)+1\big] = %
   \frac{1}{4\pi}\bigg[\big(\Lambda k\big)^2 + \left(\frac{\Lambda}{l}\right)^{\!\!2}\varepsilon_n\bigg] - t.  %
}
\end{array}
\end{equation}
If the conditions $2Lk(1-f_n(k,t))\gg 1$ and $2Lk\,f_n(k,t)\gg 1$
are fulfilled, then the distribution function turns into the usual
Fermi-Dirac distribution for every discrete level $(n)$
\begin{equation} \label{61}
\begin{array}{l}
\displaystyle{%
   f_n^{{\rm FD}}(k,t)=\Bigg[\exp\!\bigg\{\frac{1}{4\pi}\bigg[\big(\Lambda k\big)^2 + %
   \left(\frac{\Lambda}{l}\right)^{\!\!2}\varepsilon_n\bigg] - t\bigg\}+1\Bigg]^{\!-1}= %
   \bigg[e^{\frac{(\Lambda k)^2}{4\pi}-t_n} + 1\bigg]^{\!-1},
}%
\end{array}
\end{equation}
where
\begin{equation} \label{62}
\begin{array}{l}
\displaystyle{%
  t_n\equiv t - \frac{\varepsilon_n}{4\pi}\left(\frac{\Lambda}{l}\right)^{\!\!2}. %
}%
\end{array}
\end{equation}

The distribution function calculated from equation (\ref{60}) and
the Fermi-Dirac distribution (\ref{61}) are shown in Fig.\,2. The
peculiarity of the distribution function (\ref{60}) is that it
decreases from unity to zero in the finite interval of variation of
the wave number $k_1\le k\le k_2$, whereas the Fermi-Dirac function
(\ref{61}) monotonically decreases to zero in the interval $0\le k < \infty$. %
A noticeable difference between these functions takes place only
near the points $k_1$ and $k_2$. This difference has little effect
in calculations of integral quantities, so that one can use the
Fermi-Dirac distribution function (\ref{61}) for calculations in the
continual approximation.

In this approximation the total number of particles at level $(n)$
with the use of the Fermi-Dirac distribution (\ref{61}) is
\begin{equation} \label{63}
\begin{array}{l}
\displaystyle{%
  N_n=2\left(\frac{L}{\Lambda}\right)^{\!\!2}\int_0^\infty\frac{dy}{e^{y-t_n}+1}=%
      2\left(\frac{L}{\Lambda}\right)^{\!\!2}\Phi_1(t_n). %
}%
\end{array}
\end{equation}
Here the Fermi-Stoner functions at $s>0$ are defined by the formula \cite{DS,PS4} %
\begin{equation} \label{64}
\begin{array}{l}
\displaystyle{%
  \Phi_s(t)=\frac{1}{\Gamma(s)}\int_0^\infty\!\frac{z^{s-1}dz}{e^{z-t}+1}. %
}%
\end{array}
\end{equation}
Note that
\begin{equation} \label{65}
\begin{array}{c}
\displaystyle{%
  \Phi_1(t)=\ln\!\big(1+e^t\big), \quad %
  \Phi_2(t)=t\ln\!\big(1+e^t\big) - \frac{t^2}{2}+\frac{\pi^2}{12}+\int_0^t\frac{xdx}{e^x+1}, %
}\vspace{3mm}\\ %
\displaystyle{\hspace{0mm}%
  \Phi_0(t)\equiv\frac{d\Phi_1(t)}{dt}=\frac{e^t}{\big(e^t+1\big)}. %
}%

\end{array}
\end{equation}
The parameter $t=\mu/T$ is related to the total number of particles by the relation %
\begin{equation} \label{66}
\begin{array}{l}
\displaystyle{%
  N=\sum_{n=1}^\infty N_n = 2\left(\frac{L}{\Lambda}\right)^{\!\!2}\sum_{n=1}^\infty\Phi_1(t_n). %
}%
\end{array}
\end{equation}

In what follows, it is convenient to introduce the following
notations for sums with Fermi-Stoner functions:
\begin{equation} \label{67}
\begin{array}{c}
\displaystyle{%
  a_{00}\equiv\sum_{n=1}^\infty\Phi_0(t_n), \quad a_{01}\equiv\sum_{n=1}^\infty\varepsilon_n\Phi_0(t_n), \quad a_{02}\equiv\sum_{n=1}^\infty\varepsilon_n^2\Phi_0(t_n),  %
}\vspace{3mm}\\ %
\displaystyle{\hspace{0mm}%
  a_{10}\equiv\sum_{n=1}^\infty\Phi_1(t_n), \quad a_{11}\equiv\sum_{n=1}^\infty\varepsilon_n\Phi_1(t_n), \quad a_{20}\equiv\sum_{n=1}^\infty\Phi_2(t_n).  %
}%
\end{array}
\end{equation}
In these notations, in the continuum approximation, the number of
particles $N$, energy $E$, entropy $S$, thermodynamic potential
$\Omega$, surface tension $\sigma$ and induction $D$ are written as
\begin{equation} \label{68}
\begin{array}{c}
\displaystyle{%
  N=2\,\frac{A}{\Lambda^2}\,a_{10}, \quad E=2T\frac{A}{\Lambda^2}\,a_{20}+\frac{T}{2\pi}\frac{A}{l^2}\,a_{11}, %
}\vspace{3mm}\\ %
\displaystyle{\hspace{0mm}%
  S=2\,\frac{A}{\Lambda^2}\bigg[2a_{20}-ta_{10}+\frac{1}{4\pi}\left(\frac{\Lambda}{l}\right)^{\!\!2}a_{11}\bigg], %
}\vspace{3mm}\\ %
\displaystyle{\hspace{0mm}%
  \Omega=-2T\frac{A}{\Lambda^2}\,a_{20}, \quad \sigma=-\frac{2T}{\Lambda^2}\,a_{20}, \quad D=\frac{T}{3\pi F}\frac{A}{l^2}\,a_{11}. %
}%
\end{array}
\end{equation}

When calculating heat capacities in the continuum approximation, it
is more convenient to proceed not from the general formulas
(\ref{52})\,--\,(\ref{54}), but to perform a calculation on the
basis of formulas (\ref{68}). Under the condition of constant number
of particles $dN=0$, we have the following relations
\begin{equation} \label{69}
\begin{array}{c}
\displaystyle{%
  T\frac{dS}{dT}=2\left(\frac{L}{\Lambda}\right)^{\!\!2} %
  \Bigg[2a_{20}-\frac{a_{10}^2}{a_{00}}+\frac{1}{2\pi}\!\left(\frac{\Lambda}{l}\right)^{\!\!2}\!\bigg(a_{11}-\frac{a_{10}a_{01}}{a_{00}}\bigg) + %
  \frac{1}{(4\pi)^2}\!\left(\frac{\Lambda}{l}\right)^{\!\!4}\!\bigg(a_{02}-\frac{a_{01}^2}{a_{00}}\bigg)\Bigg] +%
}\vspace{2mm}\\ %
\displaystyle{%
  +\frac{2T}{\Lambda^2}\Bigg[2a_{20}-\frac{a_{10}^2}{a_{00}}+\frac{1}{4\pi} %
  \!\left(\frac{\Lambda}{l}\right)^{\!\!2}\!\bigg(a_{11}-\frac{a_{10}a_{01}}{a_{00}}\bigg)\Bigg]\frac{dA}{dT}, %
}%
\end{array}
\end{equation}\vspace{-3mm}
\begin{equation} \label{70}
\begin{array}{c}
\displaystyle{%
  d\sigma=-\frac{2}{\Lambda^2}\Bigg[2a_{20}-\frac{a_{10}^2}{a_{00}}+\frac{1}{4\pi} %
  \!\left(\frac{\Lambda}{l}\right)^{\!\!2}\!\bigg(a_{11}-\frac{a_{10}a_{01}}{a_{00}}\bigg)\Bigg]dT +  %
  \frac{2T}{\Lambda^2}\frac{a_{10}^2}{a_{00}}\frac{dA}{A}, %
}%
\end{array}
\end{equation}\vspace{-3mm}
\begin{equation} \label{71}
\begin{array}{c}
\displaystyle{%
  dD=\frac{A}{3\pi Fl^2}\Bigg[\bigg(a_{11}-\frac{a_{10}a_{01}}{a_{00}}\bigg)+\frac{1}{4\pi} %
  \!\left(\frac{\Lambda}{l}\right)^{\!\!2}\!\bigg(a_{02}-\frac{a_{01}^2}{a_{00}}\bigg)\Bigg]dT +  %
  \frac{T}{3\pi Fl^2}\bigg(a_{11}-\frac{a_{10}a_{01}}{a_{00}}\bigg)dA. %
}%
\end{array}
\end{equation}
From here we find formulas for the heat capacities:
\begin{equation} \label{72}
\begin{array}{c}
\displaystyle{%
  C_{N,F,A}=\frac{2A}{\Lambda^2} %
  \Bigg[2a_{20}-\frac{a_{10}^2}{a_{00}}+\frac{1}{2\pi}\!\left(\frac{\Lambda}{l}\right)^{\!\!2}\!\bigg(a_{11}-\frac{a_{10}a_{01}}{a_{00}}\bigg) + %
  \frac{1}{(4\pi)^2}\!\left(\frac{\Lambda}{l}\right)^{\!\!4}\!\bigg(a_{02}-\frac{a_{01}^2}{a_{00}}\bigg)\Bigg],%
}%
\end{array}
\end{equation}\vspace{-3mm}
\begin{equation} \label{73}
\begin{array}{c}
\displaystyle{%
  C_{N,F,\sigma}=C_{N,F,A}+\frac{2A}{\Lambda^2}\frac{a_{00}}{a_{10}^2} %
  \Bigg[2a_{20}-\frac{a_{10}^2}{a_{00}}+\frac{1}{4\pi} %
  \!\left(\frac{\Lambda}{l}\right)^{\!\!2}\!\bigg(a_{11}-\frac{a_{10}a_{01}}{a_{00}}\bigg)\Bigg]^{\!2},  %
}%
\end{array}
\end{equation}\vspace{-3mm}
\begin{equation} \label{74}
\begin{array}{c}
\displaystyle{%
  C_{N,F,D}=C_{N,F,A}-\frac{2A}{\Lambda^2}\Bigg[2a_{20}-\frac{a_{10}^2}{a_{00}}+\frac{1}{4\pi} %
  \!\left(\frac{\Lambda}{l}\right)^{\!\!2}\!\bigg(a_{11}-\frac{a_{10}a_{01}}{a_{00}}\bigg)\Bigg]\! %
  \Bigg[1+\frac{1}{4\pi}\!\left(\frac{\Lambda}{l}\right)^{\!\!2}\!\bigg(a_{02}-\frac{a_{01}^2}{a_{00}}\bigg)\! %
  \bigg(a_{11}-\frac{a_{10}a_{01}}{a_{00}}\bigg)^{\!\!-1}\Bigg]. %
}%
\end{array}
\end{equation}

At large surface densities $n_A\equiv N/A$, such that $n_Al^2\gg 1$,
when particles are distributed over a large number of levels, at
calculation of sums (\ref{67}) it is possible to pass from summation
to integration, assuming according to (\ref{17})
$\varepsilon_n=\left[\frac{3\pi}{2}\!\left(n-\frac{1}{4}\right)\right]^{2/3}$.
As a result, the quantities (\ref{67}) can be expressed through the functions (\ref{64}): %
\begin{equation} \label{75}
\begin{array}{c}
\displaystyle{%
  a_{00}=4\pi\left(\frac{T}{T_B}\right)^{\!\!3/2}\Phi_{3/2}(t), \quad %
  a_{01}=24\pi^2\left(\frac{T}{T_B}\right)^{\!\!5/2}\Phi_{5/2}(t), \quad %
  a_{02}=240\pi^3\left(\frac{T}{T_B}\right)^{\!\!7/2}\Phi_{7/2}(t),  %
}\vspace{3mm}\\ %
\displaystyle{\hspace{0mm}%
  a_{10}=4\pi\left(\frac{T}{T_B}\right)^{\!\!3/2}\Phi_{5/2}(t), \quad %
  a_{11}=24\pi^2\left(\frac{T}{T_B}\right)^{\!\!5/2}\Phi_{7/2}(t), \quad %
  a_{20}=4\pi\left(\frac{T}{T_B}\right)^{\!\!3/2}\Phi_{7/2}(t). %
}%
\end{array}
\end{equation}
Here the temperature $T_B$ is introduced, at which the
characteristic length $l$ becomes equal to the de Broglie thermal
wavelength $l=\Lambda_B\equiv\big(2\pi\hbar^2/mT_B\big)^{1\!/2}$. %

Taking into account (\ref{75}), we obtain expressions for the heat
capacities per one particle through the standard functions (\ref{64}): %
\begin{equation} \label{76}
\begin{array}{c}
\displaystyle{%
  c_{F,A}\equiv\frac{C_{N,F,A}}{N}=\frac{35}{4}\bigg[\frac{\Phi_{7/2}(t)}{\Phi_{5/2}(t)}-\frac{5}{7}\frac{\Phi_{5/2}(t)}{\Phi_{3/2}(t)}\bigg], %
}%
\end{array}
\end{equation}\vspace{-3mm}
\begin{equation} \label{77}
\begin{array}{c}
\displaystyle{%
  c_{F,\sigma}\equiv\frac{C_{N,F,\sigma}}{N}=\frac{49}{4}\frac{\Phi_{3/2}(t)\Phi_{7/2}(t)}{\Phi_{5/2}^2(t)} %
  \bigg[\frac{\Phi_{7/2}(t)}{\Phi_{5/2}(t)}-\frac{5}{7}\frac{\Phi_{5/2}(t)}{\Phi_{3/2}(t)}\bigg], %
}%
\end{array}
\end{equation}\vspace{-3mm}
\begin{equation} \label{78}
\begin{array}{c}
\displaystyle{%
  c_{F,D}\equiv\frac{C_{N,F,D}}{N}=-\frac{7}{2}\frac{\Phi_{7/2}(t)}{\Phi_{5/2}(t)} %
  \bigg[\frac{\Phi_{7/2}(t)}{\Phi_{5/2}(t)}-\frac{5}{7}\frac{\Phi_{5/2}(t)}{\Phi_{3/2}(t)}\bigg]\! %
  \bigg[\frac{\Phi_{7/2}(t)}{\Phi_{5/2}(t)}-\frac{\Phi_{5/2}(t)}{\Phi_{3/2}(t)}\bigg]^{\!-1}. %
}%
\end{array}
\end{equation}
These heat capacities depend on two combinations of functions $\Phi_s(t)$, namely %
\begin{equation} \label{79}
\begin{array}{c}
\displaystyle{%
  \Psi_1(t)\equiv \frac{\Phi_{7/2}(t)}{\Phi_{5/2}(t)}-\frac{5}{7}\frac{\Phi_{5/2}(t)}{\Phi_{3/2}(t)}, \quad %
  \Psi_2(t)\equiv\frac{\Phi_{5/2}^2(t)}{\Phi_{3/2}(t)\Phi_{7/2}(t)}, %
}%
\end{array}
\end{equation}
so that
\begin{equation} \label{80}
\begin{array}{c}
\displaystyle{%
  c_{F,A}=\frac{35}{4}\,\Psi_1(t), \quad %
  c_{F,\sigma}=\frac{49}{4}\frac{\Psi_1(t)}{\Psi_2(t)}, \quad %
  c_{F,D}=\frac{7}{2}\frac{\Psi_1(t)}{\big[\Psi_2(t)-1\big]}. %
}%
\end{array}
\end{equation}
Note that $\Psi_1(t)>0$, and $\Psi_2(t)>1$. %

For the surface density of the particles number $n_A\equiv
N/A=2a_{10}\big/\Lambda^2$, with account of (\ref{75}), we have
\begin{equation} \label{81}
\begin{array}{c}
\displaystyle{%
  \frac{n_Al^2}{8\pi}\left(\frac{T_B}{T}\right)^{\!\!5/2} = \Phi_{5/2}(t). %
}%
\end{array}
\end{equation}
Formulas (\ref{80}), together with (\ref{81}), parametrically define
the dependences of heat capacities on the temperature and surface density %
for the case of large surface densities $n_Al^2\gg 1$.

In the limit $t\rightarrow +\infty$, which corresponds to low
temperatures, we have the asymptotics
\begin{equation} \label{82}
\begin{array}{c}
\displaystyle{%
  c_{F,A}=c_{F,\sigma}=c_{F,D}\sim \frac{5\pi^2}{6}\frac{1}{t}. %
}%
\end{array}
\end{equation}
In this case, according to (\ref{81}),
$\displaystyle{t=\left(\frac{15}{64}\frac{n_Al^2}{\sqrt{\pi}}\right)^{\!\!2/5}\frac{T_B}{T}}$, %
so that all heat capacities at $T\rightarrow 0$, as is typical for
Fermi systems, depend linearly on temperature
\begin{equation} \label{83}
\begin{array}{c}
\displaystyle{%
  c_{F,A}=c_{F,\sigma}=c_{F,D}= \frac{5\pi^2}{6} %
  \left(\frac{64\sqrt{\pi}}{15\,n_Al^2}\right)^{\!\!2/5}\!\frac{T}{T_B}. %
}%
\end{array}
\end{equation}

Let us also consider the limit $t\rightarrow -\infty$, which
corresponds to high temperatures. We can use the approximation
$\Phi_s(t)\approx e^t$ \cite{PS4}. In this case %
$c_{F,A}\approx 5/2 - \!\big(15\big/32\sqrt{2}\big)\,e^t$, %
$c_{F,\sigma}\approx 7/2 - \!\big(35\big/32\sqrt{2}\big)\,e^t$, %
$c_{F,D}\approx 2^{7\!/2}\big/e^t$. Since, according to (\ref{81}),
$\displaystyle{e^t}\approx\frac{n_Al^2}{8\pi}\!\left(\frac{T_B}{T}\right)^{\!\!5/2}$, %
then at high temperatures we have the following dependencies for
heat capacities
\begin{equation} \label{84}
\begin{array}{c}
\displaystyle{%
  c_{F,A}=\frac{5}{2}-\frac{15}{32\sqrt{2}}\,\,\alpha\left(\frac{T_B}{T}\right)^{\!\!5/2}, %
}%
\end{array}
\end{equation}\vspace{-3mm}
\begin{equation} \label{85}
\begin{array}{c}
\displaystyle{%
  c_{F,\sigma}=\frac{7}{2}-\frac{35}{32\sqrt{2}}\,\,\alpha\left(\frac{T_B}{T}\right)^{\!\!5/2}, %
}%
\end{array}
\end{equation}\vspace{-3mm}
\begin{equation} \label{86}
\begin{array}{c}
\displaystyle{%
  c_{F,D}=\frac{8\sqrt{2}}{\alpha}\left(\frac{T}{T_B}\right)^{\!\!5/2}, %
}%
\end{array}
\end{equation}
where $\alpha\equiv n_Al^2/8\pi$. As we can see, with increasing
temperature the first two heat capacities tend to constant values
$c_{F,A}=5/2$ and $c_{F,\sigma}=7/2$, while the third heat capacity
increases as $c_{F,D}\sim T^{5/2}$.

As noted above (\ref{55}), the system under consideration is
spatially inhomogeneous in the direction of the field. Let us
calculate the density distribution in the continuum limit when
formula (\ref{63}) is valid. Then, taking into account the form of
the wave function (\ref{12}), we have
\begin{equation} \label{87}
\begin{array}{c}
\displaystyle{%
  n(z)=\frac{2}{\Lambda^2l}\sum_{n=1}^\infty\!
  \left(\frac{{\rm Ai}(\tilde{z}-\varepsilon_n)}{{\rm Ai}'(-\varepsilon_n)}\right)^{\!2}\Phi_1(t_n). %
}%
\end{array}
\end{equation}
Along with this one should consider that, according to (\ref{66}),
the parameter $t=\mu/T$ is related to the surface density by the relation %
\begin{equation} \label{88}
\begin{array}{c}
\displaystyle{%
  n_A=\frac{2}{\Lambda^2}\sum_{n=1}^\infty\Phi_1(t_n). %
}%
\end{array}
\end{equation}
Due to the boundary condition the density on the surface $n(0)=0$,
but at a small distance $\tilde{z}_m<\varepsilon_1$ the density
reaches a maximum and further decreases with increasing distance.
The maximum of the density is determined by the first maximum of
wave functions. At zero temperature, as follows from the general
formulas (\ref{87}),\,(\ref{88}), the density behavior is determined
by the relations
\begin{equation} \label{89}
\begin{array}{c}
\displaystyle{%
  n(z)=\frac{1}{2\pi l^3}\sum_{n=1}^N\!
  \left(\frac{{\rm Ai}(\tilde{z}-\varepsilon_n)}{{\rm Ai}'(-\varepsilon_n)}\right)^{\!2}\!\big(\tilde{\mu}-\varepsilon_n\big), %
}%
\end{array}
\end{equation}\vspace{-3mm}
\begin{equation} \label{90}
\begin{array}{c}
\displaystyle{%
  n_A=\frac{1}{2\pi l^2}\sum_{n=1}^N\!\big(\tilde{\mu}-\varepsilon_n\big), %
}%
\end{array}
\end{equation}
where $\tilde{\mu}\equiv 4\pi\mu/T_B$. This dependence for three
values of the dimensionless density $n_Al^2$ is shown in Fig.\,3({\it a}). %

\begin{figure}[h!]
\vspace{-2mm}  \hspace{0mm}
\includegraphics[width = 14.9cm]{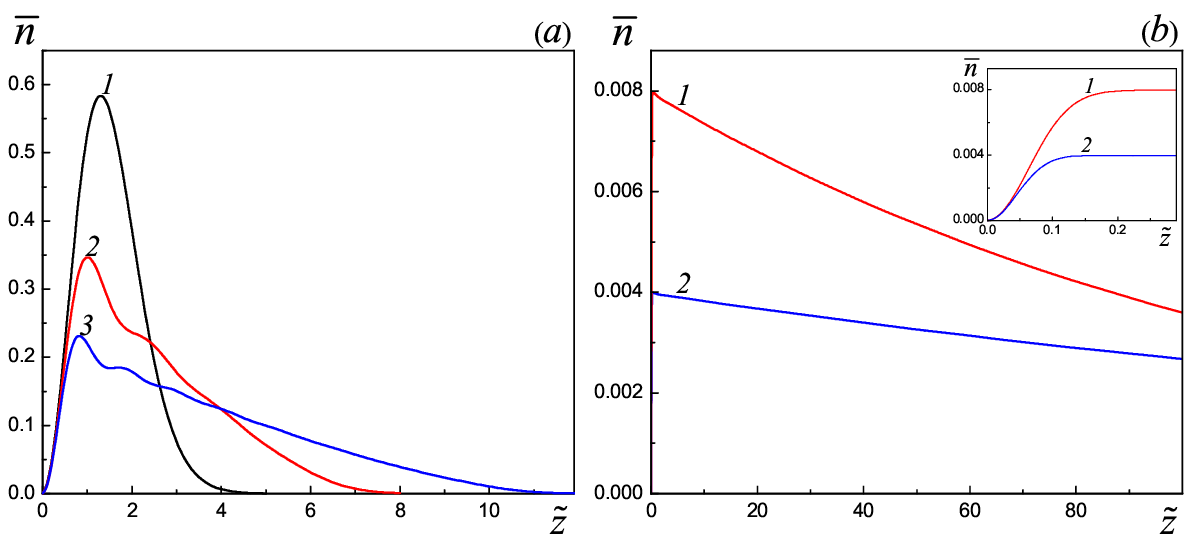} % 1.0\columnwidth
\vspace{-4mm} %
\caption{\label{fig03} %\hspace{10mm} %
({\it a}) The dependence $\overline{n}(\tilde{z})\equiv
l\,n(\tilde{z})/n_A$, calculated by (\ref{89}),\,(\ref{90}), at
$T=0$ and values of $n_Al^2$, corresponding to
$\tilde{\mu}=\varepsilon_k+0.5(\varepsilon_{k+1}-\varepsilon_k)$:
({\it 1}) 0.14 for $k=1$, ({\it 2}) 1.7 for $k=4$, ({\it 3}) 5.6 for $k=8$. %
({\it b}) The dependence $\overline{n}(\tilde{z})$ at temperatures
$T/T_B$: ({\it 1}) 10, ({\it 2}) 20. The inset shows the behavior
of $\overline{n}(\tilde{z})$ near zero. %
}%
\end{figure}

In the limit of high temperatures at $t<0$ and $T/T_B\gg 1$  we have %
\begin{equation} \label{91}
\begin{array}{c}
\displaystyle{%
  n(z)=\frac{n_A}{4\pi l}\left(\frac{T}{T_B}\right)^{\!\!-3\!/2} %
  \sum_{n=1}^\infty\!\left(\frac{{\rm Ai}(\tilde{z}-\varepsilon_n)}{{\rm Ai}'(-\varepsilon_n)}\right)^{\!2}
  e^{-\frac{\varepsilon_n}{4\pi}\frac{T_B}{T}} . %
}%
\end{array}
\end{equation}
This dependence is shown in Fig.\,3({\it b}). Here as well the
density increases rapidly to a maximum value and then decreases
rather slowly with distance. %
Note that $\overline{n}(\tilde{z})$ does not depend on the density
parameter $n_Al^2$.

Let us apply the obtained general relations to the gravitational and
electric fields, limiting ourselves in this paper to the continual
approximation.

\section{Gravitational and electric fields} %
In a gravitational field $F=mg$, the characteristic length for a
particle with an electron mass $l_e=8.8\cdot10^{-2}$\,cm, and for a
neutron $l_n=5.85\cdot10^{-4}$\,cm. The corresponding temperatures
$T_B$, defined by the relation %
$l=\Lambda_B\equiv\big(2\pi\hbar^2/mT_B\big)^{1\!/2}$, %
are equal to $T_{Be}=0.7\cdot 10^{-8}$\,K and $T_{Bn}=0.88\cdot
10^{-7}$\,K for an electron and a neutron, respectively. At present,
the minimum temperatures achievable in experiments are $T\sim
10^{-3}$\,K, so that at all temperatures the gravitational field
should be described classically, when the heat capacities are %
$c_{F,A}=5/2$ and $c_{F,\sigma}=7/2$, while $c_{F,D}\sim T^{5/2}$. %

Let us consider a gas of electrons over a positively charged surface
with the charge density $\sigma_q$, assuming that the system is
electrically neutral. It should be noted that electrons over the
surface of liquid helium in the electric field have been studied in
detail both experimentally and theoretically \cite{ShM,MK}. In this
case, the magnitude of the electric field intensity
$E=2\pi\sigma_q$, and the force acting on an electron $F=|e|E$. The
difference between this case and the case of a system in a
gravitational field, where the magnitude of the field and the
density of particles on the surface are independent, is that the
neutrality condition specifies the relationship between the
magnitude of the electric field and the density of electrons.

Let $l_*$ be the characteristic length (\ref{05}) at the field
intensity $E_*$. Then the dependence of $l$ on the intensity has the form %
\begin{equation} \label{92}
\begin{array}{c}
\displaystyle{%
  \frac{l}{l_*}=\left(\frac{E}{E_*}\right)^{\!\!-1/3}. %
}%
\end{array}
\end{equation}
Thus, at the intensity $E_*=100$\,V/cm  we have $l_*=1.56\cdot
10^{-6}$\,cm, and the temperature determined by the condition
$l_*=\big(2\pi\hbar^2/mT_{B*}\big)^{1\!/2}$ is equal to
$T_{B*}=22.8$\,K. Then $T_B$ increases with increasing field as
\begin{equation} \label{93}
\begin{array}{c}
\displaystyle{%
  \frac{T_B}{T_{B*}}=\left(\frac{E}{E_*}\right)^{\!\!2/3}. %
}%
\end{array}
\end{equation}
In a neutral system, the surface charge density of the positively
charged surface must be compensated by the surface charge density of electrons %
\begin{equation} \label{94}
\begin{array}{c}
\displaystyle{%
  \sigma_q=\frac{E}{2\pi}=|e|n_A. %
}%
\end{array}
\end{equation}
Thus, at $E_*=100$\,V/cm the surface density $n_{A*}=1.1\cdot
10^8$\,cm$^{-2}$, and it is proportional to the field intensity
\begin{equation} \label{95}
\begin{array}{c}
\displaystyle{%
  \frac{n_{A}}{n_{A*}}=\frac{E}{E_*}. %
}%
\end{array}
\end{equation}
The dimensionless surface density slowly increases with the field
intensity $n_Al^2=n_{A*}l_*^2\,\big(E/E_*\big)^{\!1/3}$, where
$n_{A*}l_*^2=2.68\cdot 10^{-4}$. It follows that for all reasonable
values of field intensities and at low temperatures, calculations
should be carried out using formulas that take into account the
discreteness of the levels. %
In Fig.\,4, the dependence of $n_{A,n}/n_A\equiv N_n/N$ on
temperature for some lower levels at the field intensity $E=E_*$ is
presented. At low surface density
near zero temperature, all particles are at the bottom level, so that $n_{A,1}/n_A=1$. %
With increasing temperature there begin transitions of particles
from the first to higher levels $n>1$. %
%At $T/T_B>1$ it is possible to use formulas (\ref{75}) in calculations. %
In the limit of high temperatures $T/T_B\gg1$, formulas
(\ref{75}),(\ref{81}) are valid for arbitrary surface densities $n_Al^2$. %
\begin{figure}[h!]
\vspace{-0mm}  \hspace{0mm}
\includegraphics[width = 7.6cm]{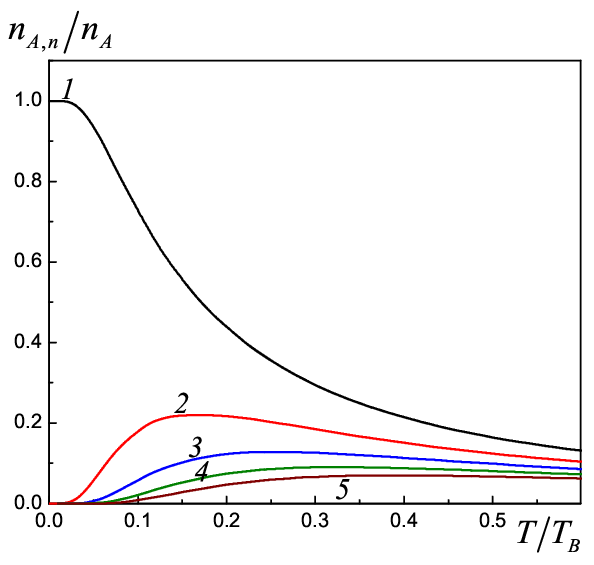} % 1.0\columnwidth
\vspace{-4mm} %
\caption{\label{fig04} %\hspace{10mm} %
The dependencies of populations $n_{A,n}/n_{A}$ on temperature for
levels $n=1\,\div\,5$ at the field intensity $E=E_*$. The filling of
the second level begins at $T/T_{B}\approx 0.02$. The normalization
condition $\sum_{n=1}^\infty\!\frac{n_{A,n}}{n_{A}}=1$.
}%
\end{figure}

Taking into account formulas (\ref{92})\,--\,(\ref{95}), from
(\ref{66}) we find the relationship of the parameter $t$ with the
field intensity and temperature
\begin{equation} \label{96}
\begin{array}{c}
\displaystyle{%
  n_{A*}l_*^2=2\,\frac{T}{T_{B*}}\frac{E_*}{E}\sum_{n=1}^\infty %
  \Phi_1\bigg(t-\frac{\varepsilon_n}{4\pi}\frac{T_{B*}}{T}\left(\frac{E}{E_*}\right)^{\!\!2/3}\bigg). %
}%
\end{array}
\end{equation}
Taking into account (\ref{96}), we can construct the dependences of
heat capacities (\ref{72})\,--\,(\ref{74}) per one particle on
temperature at a fixed value of the field intensity (Fig.\,5) and
the dependences of heat capacities on the field intensity at a fixed
temperature (Fig.\,6).

\begin{figure}[h!]
\vspace{-2mm}  \hspace{0mm}
\includegraphics[width = 17.2cm]{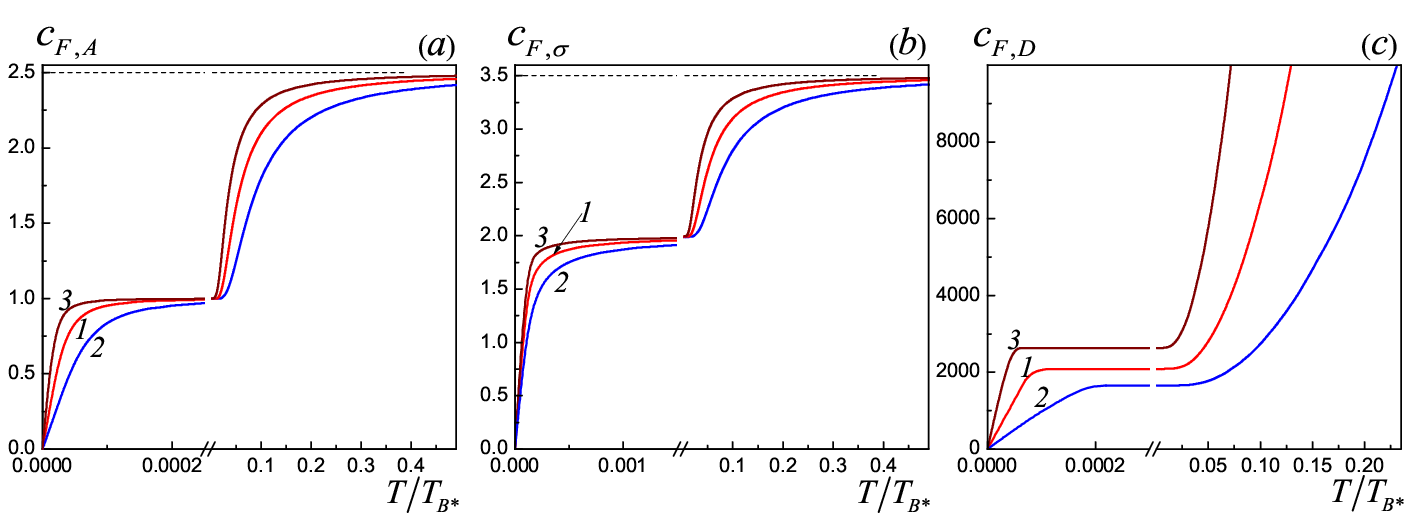} % 1.0\columnwidth
\vspace{-4mm} %
\caption{\label{fig05} %\hspace{10mm} %
The dependencies of heat capacities ({\it a}) $c_{F,A}$, ({\it b})
$c_{F,\sigma}$, ({\it c}) $c_{F,D}$ on temperature %
at $n_{A*}l_*^2$ and intensities of the electric field $E/E_*$:
({\it 1}) 1.0, ({\it 2}) 2.0, ({\it 3}) 0.5. %
Low-temperature and high-temperature regions are shown. %
}%
\end{figure}
\begin{figure}[h!]
\vspace{-3mm}  \hspace{0mm}
\includegraphics[width = 17.3cm]{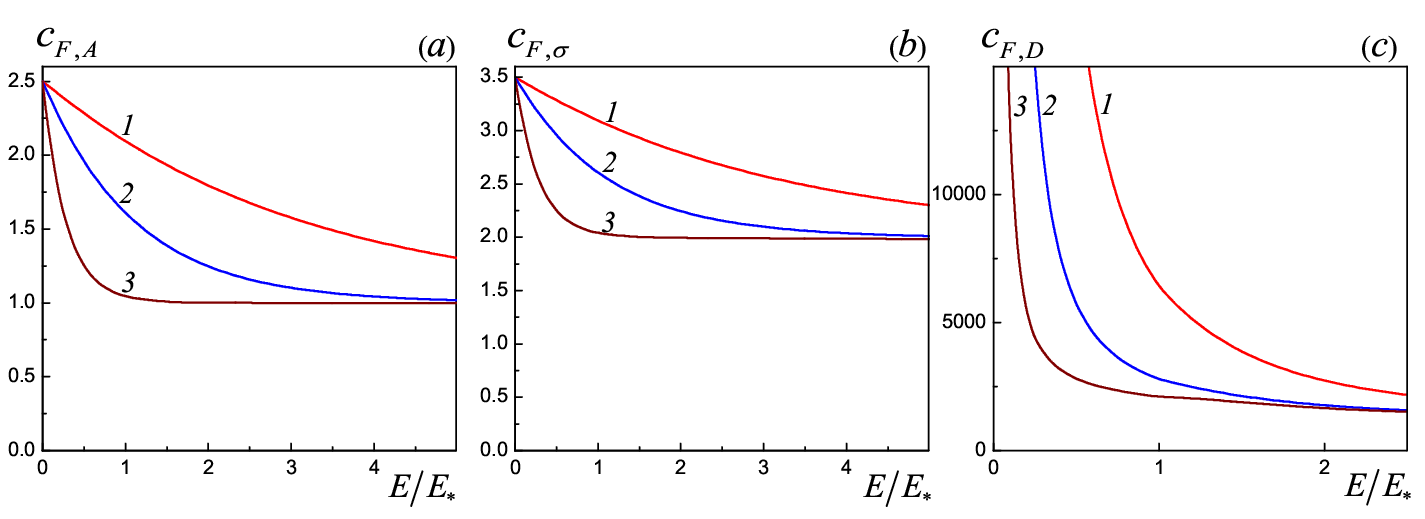} % 1.0\columnwidth
\vspace{-4mm} %
\caption{\label{fig06} %\hspace{10mm} %
The dependencies of heat capacities ({\it a}) $c_{F,A}$, ({\it b})
$c_{F,\sigma}$, ({\it c}) $c_{F,D}$ on the field intensity %
at $n_{A*}l_*^2$ and fixed temperatures $T/T_{B*}$: ({\it 1}) 0.1,
({\it 2}) 0.05, ({\it 3}) 0.02.
}%
\end{figure}

If at zero temperature the number of filled levels is greater than
one $n>1$ with $\varepsilon_n\le \tilde{\mu}\le \varepsilon_{n+1}$,
then at $T\rightarrow 0$ all heat capacities depend equally linearly
on temperature
\begin{equation} \label{97}
\begin{array}{c}
\displaystyle{%
  c_{F,A}\sim c_{F,\sigma}\sim c_{F,D}\sim \frac{2\pi^2}{3}\frac{n}{n_Al^2}\frac{T}{T_B}. %
}%
\end{array}
\end{equation}
The case, when at $T=0$ all particles are at the lower level $n=1$,
with $\varepsilon_1\le \tilde{\mu}\le \varepsilon_{2}$, is special. %
Here, for the heat capacities $c_{F,A}$ and $c_{F,\sigma}$ the
formulas (\ref{97}) remain valid with $n=1$, while in the third heat
capacity $c_{F,D}$ the coefficient in a linear dependence changes:
\begin{equation} \label{98}
\begin{array}{c}
\displaystyle{%
  c_{F,D}\sim \frac{\pi}{3}\frac{\big(\varepsilon_2-\varepsilon_1\big)}{\big(n_Al^2\big)^2}\frac{T}{T_B}. %
}%
\end{array}
\end{equation}
With an increase in temperature, the heat capacities initially reach
a plateau, and at the temperature at which transitions from the
first to higher levels arise (see Fig.\,4) there begins further
growth of the heat capacities (Fig.\,5). In the limit of high
temperatures, the heat capacities are described by formulas (\ref{84})\,--\,(\ref{86}). %
%(being valid also for the case $n_Al^2\ll 1$)

The dependences of heat capacities on the field intensity at fixed
temperature are shown in Fig.\,6. The heat capacities decrease with
increasing the field intensity and at high intensities they approach
their values on the plateaus (Fig.\,5). This corresponds to the
phenomenon that all particles in a strong field accumulate at the
lower level.

\section{Conclusion} %\vspace{-2mm}%%
A statistical approach to the description of the thermodynamic
properties of the Fermi particle system over a flat surface in a
uniform external field is proposed. At that the density of the
number of fermions per unit surface is assumed to be arbitrary and
can also be small. General formulas for the heat capacities at fixed
surface area, surface tension and induction are obtained. A
continuum approximation is considered, in which the surface area is
assumed to be large, so that the motion in the plane is
characterized by a two-dimensional wave vector. In this case, near
zero temperature, the heat capacities are proportional to
temperature. At high temperatures, the heat capacity at constant
induction increases as $c_{F,D}\sim T^{5/2}$, and the other two heat
capacities tend to constant values $c_{F,A}=5/2$ and
$c_{F,\sigma}=7/2$. The fermion density distribution over the
surface is found. The cases of gravitational and electric fields are
considered. The dependences of heat capacities in the electric field
on the temperature and field intensity are obtained.

%\vspace{4mm}


\newpage
\begin{thebibliography}{99}
\bibitem{LL}
L.D.\,Landau, E.M.\,Lifshitz, {\it Statistical physics}: Vol.\,5
(Part\,1), Pergamon Press, Oxford (1980).
\bibitem{Hill}
T.L.\,Hill, {\it Thermodynamics of small systems} (Part 1),
W.A.\,Benjamin, New York (1963).
\bibitem{PS1}
Yu.M.\,Poluektov and A.A.\,Soroka, {\it Quantum distribution
functions in systems with an arbitrary number of particles}, %
{\it Nanosystems, Nanomaterials, Nanotechnologies} \textbf{23}, 1 (2025). %
https:/\!/doi.org/10.15407/nnn.23.01.001 \,(arXiv:2311.03003v2\,[quant-ph]) %
\bibitem{PS2}
Yu.M.\,Poluektov and A.A.\,Soroka, {\it On the thermodynamics of
two-level Fermi and Bose nanosystems}, {\it Opt. Quant. Electron.} \textbf{56}, 1349 (2024). %
https:/\!/doi.org/10.1007/s11082-024-07266-x \,(arXiv:2405.02427\,[quant-ph]) %
\bibitem{PS3}
Yu.M.\,Poluektov and A.A.\,Soroka, {\it Thermodynamics of the Fermi
gas in a cubic cavity of an arbitrary volume}, {\it Eur. Phys. J.
B}\,\,\textbf{98}, 116 (2025). %
https:/\!/doi.org/10.1140/epjb/s10051-025-00967-6 \,(arXiv:2408.03667\,[quant-ph]) %
\bibitem{PS3b}
Yu.M.\,Poluektov and A.A.\,Soroka, {\it Thermodynamics of the Fermi
gas in a sphere of arbitrary radius}, {\it Metallophysics and
Advanced Technologies} \textbf{47}, 907 (2025). https:/\!/doi.org/10.15407/mfint.47.09.0907 %
\bibitem{GDMD}
J.A.\,Gil-Corrales, C.A.\,Dagua-Conda, M.E.\,Mora-Ramos,
A.L.\,Morales, C.A.\,Duque, {\it Shape and size effects on
electronic thermodynamics in nanoscopic quantum dots}, {\it Physica
E}\,\,\textbf{170}, 116228 (2025). https:/\!/doi.org/10.1016/j.physe.2025.116228 %
\bibitem{AS}
{\it Handbook of Mathematical Functions with Formulas, Graphs, and
Mathematical Tables}, M.\,Abramowitz, I.\,Stegun (eds.), %
{\it Applied Mathematics Series} \textbf{55}, National Bureau of Standards (1964). %
\bibitem{DS}
J.\,McDougall, E.C.\,Stoner, {\it The computation of Fermi-Dirac
functions}, {\it Phil. Trans. Roy. Soc. A}\,\,\textbf{237}, 67 (1938). %
https:/\!/doi.org/10.1098/rsta.1938.0004
\bibitem{PS4}
Yu.M.\,Poluektov and A.A.\,Soroka, {\it Thermodynamic functions of
the Fermi gas at arbitrary temperatures} (2024).
arXiv:2410.16332\,[cond-mat.quant-gas]
\bibitem{ShM}
V.B.\,Shikin, Y.P.\,Monarkha, {\it Two-dimensional charged systems
in helium}, Nauka, Moscow (1989).
\bibitem{MK}
Y.\,Monarkha, K.\,Kono, {\it Two-dimensional Coulomb liquids and
solids}, Springer, Berlin (2004).
\end{thebibliography}
\end{document}